\documentclass[aps,prl,twocolumn,showpacs,amsmath,amssymb,floatfix,superscriptaddress]{revtex4-1}

\usepackage{graphicx}
\usepackage{dcolumn}
\usepackage{bm}
\usepackage{color}
\usepackage{graphicx}
\usepackage{epstopdf}
\usepackage{epsfig}
\usepackage[colorlinks=true, citecolor=blue, anchorcolor=green]{hyperref}

\makeatletter
\begin{document}
\include{graphic}
\preprint{APS/123-QED}

\title{Ultrasensitive Magnetometer based on Cusp Points of the Photon-Magnon Synchronization Mode}

\author{Xinlin Mi}
\affiliation{School of Physics and State Key Laboratory of Crystal Materials, Shandong University, No. 27 Shandanan Road, Jinan, 250100 China}

\author{Jinwei Rao} \email{raojw@sdu.edu.cn} 
\affiliation{School of Physics and State Key Laboratory of Crystal Materials, Shandong University, No. 27 Shandanan Road, Jinan, 250100 China}

\author{Lijun Yan} 
\affiliation{School of Physics and State Key Laboratory of Crystal Materials, Shandong University, No. 27 Shandanan Road, Jinan, 250100 China}

\author{Xudong Wang} 
\affiliation{School of Physics and State Key Laboratory of Crystal Materials, Shandong University, No. 27 Shandanan Road, Jinan, 250100 China}

\author{Bingbing~Lyu}
\affiliation{School of Integrated Circuit, Shandong University, 27 Shandanan Road, Jinan, 250100 China}

\author{Bimu~Yao}
\affiliation{State Key Laboratory of Infrared Physics, Shanghai Institute of Technical Physics, Chinese Academy of Sciences, Shanghai 200083, China}
\affiliation{School of Physical Science and Technology, ShanghaiTech University, Shanghai 201210, China}

\author{Shishen Yan}
\affiliation{School of Physics and State Key Laboratory of Crystal Materials, Shandong University, No. 27 Shandanan Road, Jinan, 250100 China}

\author{Lihui Bai} \email{lhbai@sdu.edu.cn;}
\affiliation{School of Physics and State Key Laboratory of Crystal Materials, Shandong University, No. 27 Shandanan Road, Jinan, 250100 China}

\begin{abstract}

Ultrasensitive magnetometers based on spin resonances have led to remarkable achievements. However, the gyromagnetic ratios of these spin resonances that determine the responsivity of magnetometers to weak magnetic fields are inherently constrained by the Land$\acute{e}$ g-factor of particles, such as the electron, with a constant gyromagnetic ratio of $\gamma_e=2\pi\times28$ GHz/T. Here, we demonstrate an ultrasensitive magnetometer based on the cusp point (CP) of photon-magnon synchronization modes (PMSMs). The PMSM's gyromagnetic ratio at the CP is enhanced to $37\gamma_e$ and further amplified to $236\gamma_e$ by utilizing the sixth-order oscillating mode of the PMSM. Moreover, the emission linewidth of the PMSM can be reduced to 0.06 Hz, resulting in excellent sensitivity to weak magnetic fields. These outstanding properties position our magnetometer to potentially achieve superior sensitivity to conventional magnetometers. Our work introduces a cost-effective prototype for the next generation of magnetometry, and may advance scientific research and technologies that rely on ultrasensitive magnetic field detection. 

\end{abstract}
\maketitle

\begin{figure*}
\begin{center}
\epsfig{file=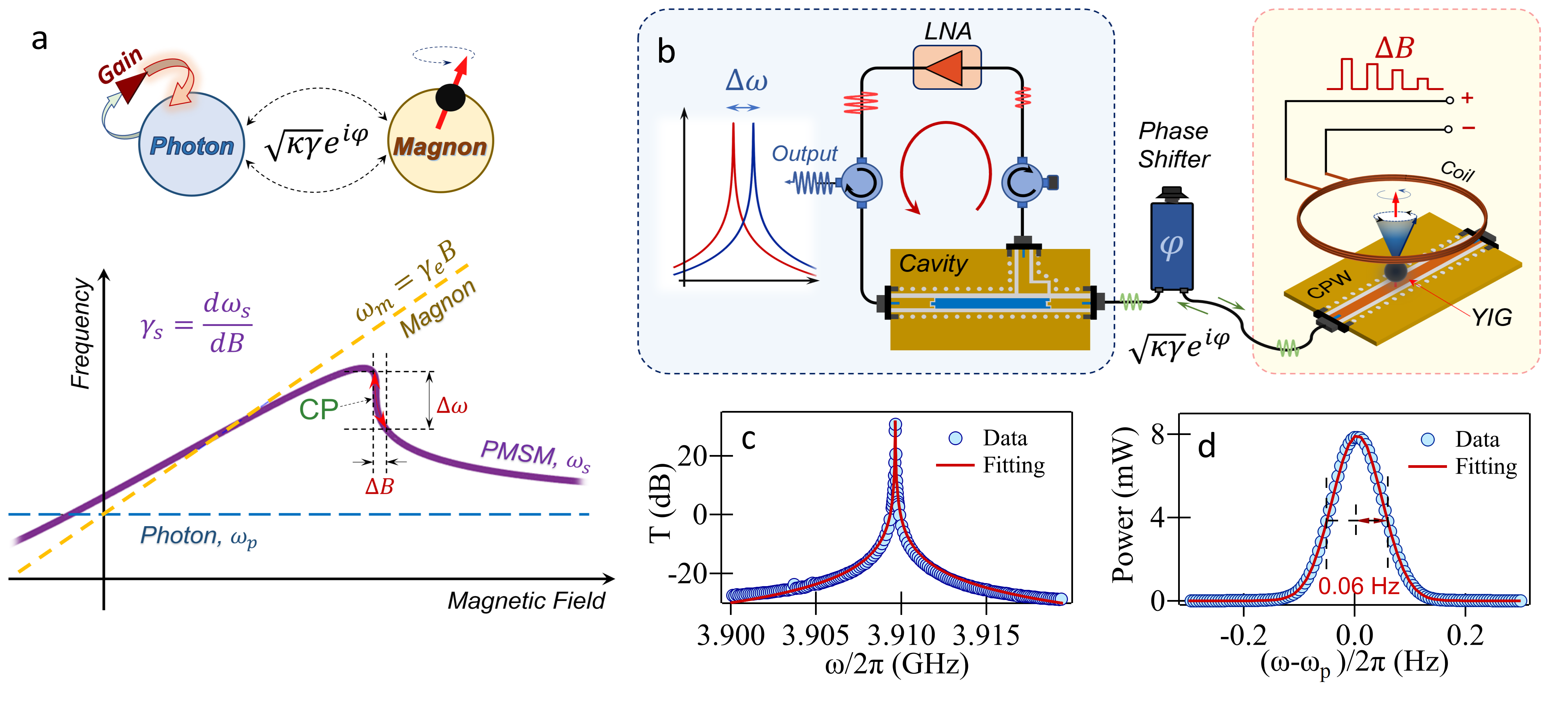, width=17cm} \caption{(a), Top: Schematic of the coupling effect between an active photon mode and a magnon mode. Their coupling strength $\sqrt{\kappa\gamma}e^{i\varphi}$ is tunable by the propagation phase of traveling waves ($\varphi$). Bottom: PMSM's frequency as a function of the magnetic field. At the CP, the PMSM's gyromagnetic ratio ($\gamma_s$) is significantly enhanced. (b), Sketch of our magnetometer. The phase shifter is used to tune the photon-magnon coupling. (c), Transmission spectrum of the fundamental photon mode of the active cavity. (d), Emission spectrum of the fundamental photon mode after locked by a weak microwave. Spectra in (c) and (d) are respectively fitted by the transmission formula and Gaussian curve. }
\label{F1}
\end{center}
\end{figure*}

Due to the significant applications of highly sensitive magnetometers in magnetoencephalography \cite{boto2018moving,brookes2022magnetoencephalography,tierney2019optically}, dark matter detection \cite{bloch2022new,jiang2021search,gramolin2021search,afach2021search}, mineral exploration \cite{liu2018nonlinear,stolz2022squids}, and various other fields \cite{carletta2020magnetometer,walter2020high,fassbinder2015seeing,bennett2021precision,alberto2018optical}, several important techniques have been developed in recent decades, such as superconducting quantum interference devices \cite{granata2016nano,vasyukov2013scanning,de2018charge}, nitrogen-vacancy centers in diamonds \cite{fescenko2020diamond,barry2020sensitivity,casola2018probing}, optomechanical systems \cite{forstner2012cavity, xu2024subpicotesla} and alkali-metal atom gases \cite{allred2002high,dang2010ultrahigh,meng2023machine, qin2024new}. Among these methods, atomic magnetometry based on alkali-metal atom gases stands out for its astonishing sensitivity. Its core mechanism involves heating dense alkali-metal atoms into the spin-exchange relaxation free state \cite{anderson1963study, happer1973spin, ledbetter2008spin}. This special spin state features an ultranarrow linewidth at several hundred hertz, and its frequency varies with the magnetic field according to a gyromagnetic ratio of $\gamma_e=2\pi\times28$ GHz/T. These two parameters determine the frequency resolution limit and the responsivity to the magnetic field of an atomic magnetometer, and they jointly determine the sensitivity of atomic magnetometry. The resonance linewidth of the spin-exchange relaxation free state can potentially be further reduced by optimizing the chamber design or increasing the atom density; however, its gyromagnetic ratio is fixed by the electron's Land$\acute{e}$ g-factor \cite{scott1962review}. This physical limitation is not unique to atomic magnetometers but is widely shared by other magnetometers based on magnetic or spin resonances.

The recently discovered photon-magnon synchronization mode (PMSM) in an active cavity magnonic system \cite{rao2023meterscale} offers a promising way to bypass this physical limitation. Generally, a microwave photon mode in an active cavity is self-sustained with a theoretically zero resonance linewidth, providing ultra-high frequency resolution but being insensitive to magnetic fields. In contrast, a magnon mode responds to magnetic fields but suffers from its broad resonance linewidth. The PMSM produced by strong photon-magnon coupling \cite{tabuchi2015coherent, wang2024enhancement, rao2023unveiling, xu2024slow, wang2018bistability, li2019strong, yuan2022quantum} combines the complementary advantages of both the active photon mode and the magnon mode, thereby achieving ultra-narrow linewidth and high magnetic responsivity \cite{rao2023meterscale}. Moreover, the PMSM's hysteresis behavior implies the existence of cusp points (CPs) \cite{strogatz2018nonlinear}, which arise from the coalescence of saddle-node bifurcations (SNBs) \cite{strogatz2018nonlinear,kuznetsov1998elements} and can be used to greatly improve the PMSM’s gyromagnetic ratio. Consequently, even a small variation in the magnetic field can cause a significant frequency shift in the PMSM (Fig. \ref{F1} (a)). Due to these advantages, we realize that the PMSM at CPs holds great promise for ultra-high sensitivity in detecting weak magnetic fields. However, this concept has not been explored yet.

In this work, we construct a magnetometer based on the CP of PMSMs and demonstrate its capability in sensitive magnetic detection. To produce the PMSM, we fabricate an active cavity and couple its fundamental photon mode to the magnon mode in a yttrium iron garnet (YIG) sphere via traveling waves. As we finely tune the photon-magnon coupling by adjusting the propagation phase of traveling waves, the PMSM transitions between bistability and unimodality. At the transition boundary, a CP occurs, where the PMSM's gyromagnetic ratio is significantly improved to $37\gamma_e$ in our experiment. Additionally, we find that the frequency multiplication in the active cavity works as a ``vernier caliper" in the frequency domain, so that a small frequency shift of the PMSM can be amplified into a significant shift by its high-order oscillating modes. Here, we use the sixth-order oscillating mode and improve the PMSM's gyromagnetic ratio at the CP by approximately six times, resulting in a recorded value of $236\gamma_e$. It means that, compared to electron spin resonances, the PMSM at the CP exhibits a frequency shift 236 times larger for the same variation in the magnetic field. Our magnetometer operates at room temperature and does not require optical pumping or readout. All its components can be integrated onto a printed circuit board. Although its current sensitivity is 3.64 fT/$\sqrt{\mathrm{Hz}}$ and still needs to be further enhanced through noise suppression, our magnetometer provides a promising and cost-effective prototype for the next generation of magnetometry.

\begin{figure*}
\begin{center}
\epsfig{file=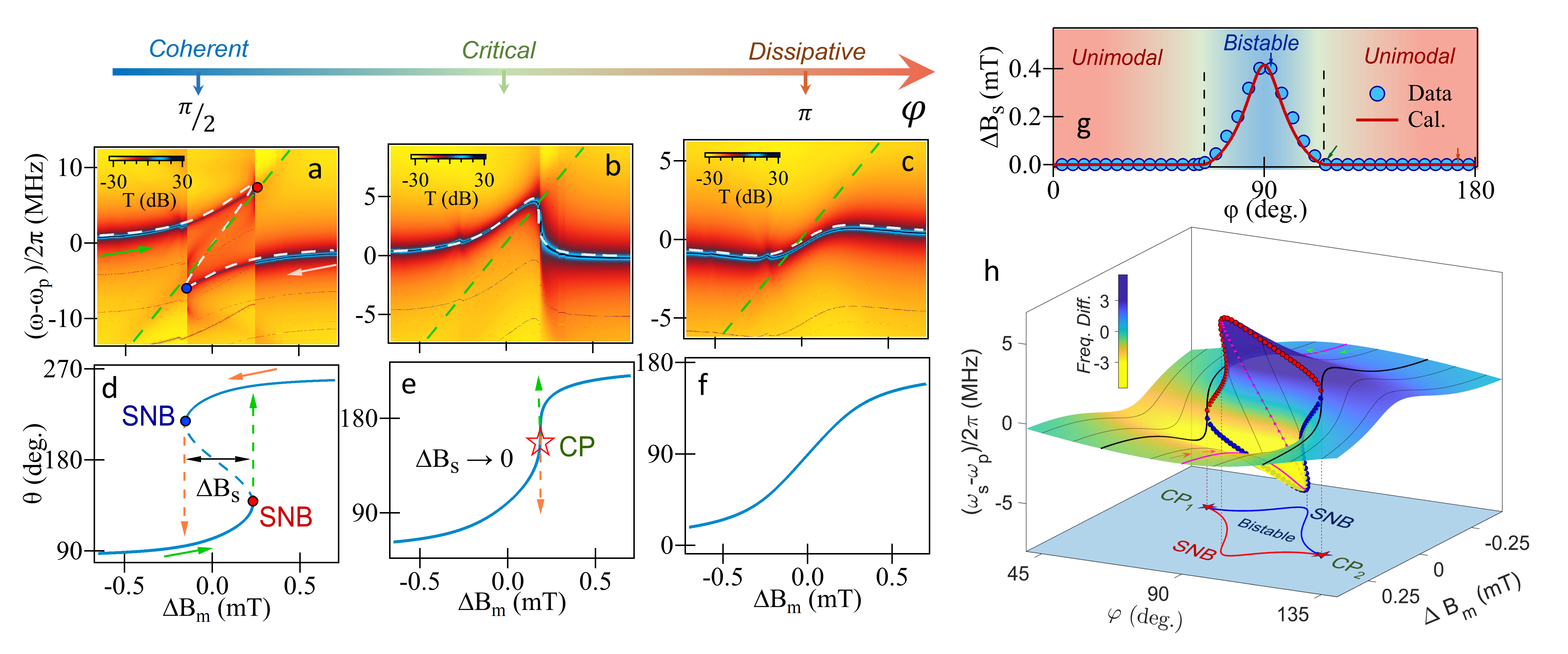, width=18cm}\caption{(a)-(c), Transmission spectra of our magnetometer measured at different magnetic fields, when the photon-magnon coupling is coherent ($\varphi\approx 93^\circ$), critical ($\varphi\approx 116^\circ$) and dissipative ($\varphi\approx 173^\circ$), respectively. The green dashed lines represent the magnon frequency, while the white dashed lines show the calculated frequency of the PMSM ($\omega_s$) using Eq. (\ref{PMSM}). The transmission spectra in (a) is the average of two measurements taken with the magnetic field swept upward and downward. (d)-(f), Phase difference ($\theta$) between the photon and magnon modes calculated from Eq. (\ref{PD}). Dashed arrows indicate the frequency jump at the SNBs (marked by blue and red circles). Red star in (e) marks the CP where two SNBs coalesce. (g), Field difference between two SNBs ($\Delta B_s$) as a function of $\varphi$. Red solid line is numerically calculated from Eq. (\ref{PD}). Vertical dashed lines indicate the transition boundaries between unimodality and bistability, where CPs occur. (h), Evolution of $\omega_s$ plotted on a two-dimensional parameter space defined by $\varphi$ and $\Delta B_m$. Red and blue circles are SNBs. CPs occur when SNBs coalesce. Black solid lines are the calculated PMSM's frequencies when CPs exist. }
\label{F2}
\end{center}
\end{figure*}

Our magnetometer consists of three main parts: an active cavity, a connecting part and a magnetic-field probe (Fig. \ref{F1} (b)). The active cavity contains a passive planar cavity and a feedback gain loop. Compared to previous work \cite{rao2023meterscale}, this active cavity shows better frequency stability. The passive cavity is a three-port device. One port is connected to the magnetic-field probe to enable the strong photon-magnon coupling. Two other ports are connected to an amplifier to form a feedback gain loop \cite{jones2014design, zarifi2015high}. In addition to providing a gain, the feedback loop can also function as a self-injection-locked loop that can stabilize the cavity oscillation. The connecting part contains a coaxial cable ($\sim 0.5$ m) and a mechanical phase shifter. By adjusting the propagation phase of traveling waves ($\varphi$) using the phase shifter, we can precisely control the photon-magnon coupling state \cite{rao2023meterscale} and hence produce the CPs. In the magnetic-field probe, a YIG sphere with a diameter of 1 mm is mounted on a coplanar waveguide (CPW). Under the CPW, a permanent magnet offers a stable bias magnetic field ($B$). Changing the separation between the permanent magnet and the CPW can tune the magnon mode frequency ($\omega_m=\gamma_eB$). Above the CPW and YIG sphere, a solenoid coil is used to fine-tune the bias magnetic field. 

The cavity photon mode can be modeled by a Hamiltonian $\mathcal{H}_a=\hbar[\omega_p-i(\kappa_t+G)]\hat{a}^\dag\hat{a}$, which includes a saturable gain $G=-N\left[1-i\sqrt{\varepsilon/2}\hat{a}-\varepsilon|\hat{a}|^2\right]$, where $\omega_p$, $\kappa_t$, $\hat{a}$, $N$ and $\varepsilon$ represent the mode frequency, total damping rate, photon operator, linear negative damping and saturation factor of the cavity mode, respectively. After turning on the amplifier, as long as $N>\kappa_t$, the cavity mode evolves into a self-sustained oscillation. The dynamic damping rate in the gain, i.e., $iN\sqrt{\varepsilon/2}\hat{a}$, leads to the frequency multiplication of the cavity oscillation, which can be leveraged to enhance the sensitivity of our magnetometer. The cavity's oscillation can be characterized by either its transmission spectrum or emission spectrum. Figure \ref{F1} (c) shows a transmission spectrum of the fundamental mode of the cavity at $\omega_p/2\pi=3.91$ GHz. The cavity emission spectrum is measured using a signal analyzer. Its emission linewidth is only 84 Hz (half linewidth at half maximum) and can be further reduced by locking the cavity with a weak microwave signal. The injection-locked emission spectrum (Fig. \ref{F1} (d)) is fitted with a Gaussian curve. The obtained emission linewidth is 0.06 Hz, which determines the frequency resolution limit of our magnetometer.

Via the mediation of traveling waves in the connecting part, the cavity photon mode can strongly couple to the magnon mode in the YIG sphere with a strength $\sqrt{\kappa\gamma}e^{i\varphi}$, where $\kappa$ and $\gamma$ respectively represent the radiation rates of the photon and magnon modes into the traveling waves. This indirect coupling is a combination of coherent coupling with a strength $-\sqrt{\kappa\gamma}\sin\varphi$ and dissipative coupling with a strength $i\sqrt{\kappa\gamma}\cos\varphi$. Tuning $\varphi$ can cause a transition in the photon-magnon coupling between these two states. At the transition boundary, the bistability of the PMSM disappears, producing a CP.

Figures \ref{F2} (a)-(c) show the transmission variation of our magnetometer with the magnetic field measured at three different coupling states. When $\varphi= 93^\circ$, the photon-magnon coupling is approximately coherent. The average of the transmission spectra measured by sweeping the magnetic field upward and downward is plotted in Fig. \ref{F2} (a). With different sweeping histories, the PMSM's frequency ($\omega_s$) exhibits bistable behavior. Two frequency jump points occur at $\Delta B_m=0.228$ mT and $-0.156$ mT, where $\Delta B_m=B-\omega_p/\gamma_e$. As we further increase $\varphi$, the photon-magnon coupling transitions towards the dissipative state (Fig. \ref{F2} (c)). The PMSM's frequency becomes unimodal and displays a central symmetry about the zero detuning, i.e., $(\omega,\Delta B_m)=(\omega_p,0)$. Between these two coupling states, there exists a critical state ($\varphi=116^\circ$ in our system), where the PMSM's bistability just disappears and two jump points coalesce. Consequently, a sharp transition in $\omega_s$ occurs near $\Delta B_m=0.187$ mT.

\begin{figure} [htp]
\begin{center}
\epsfig{file=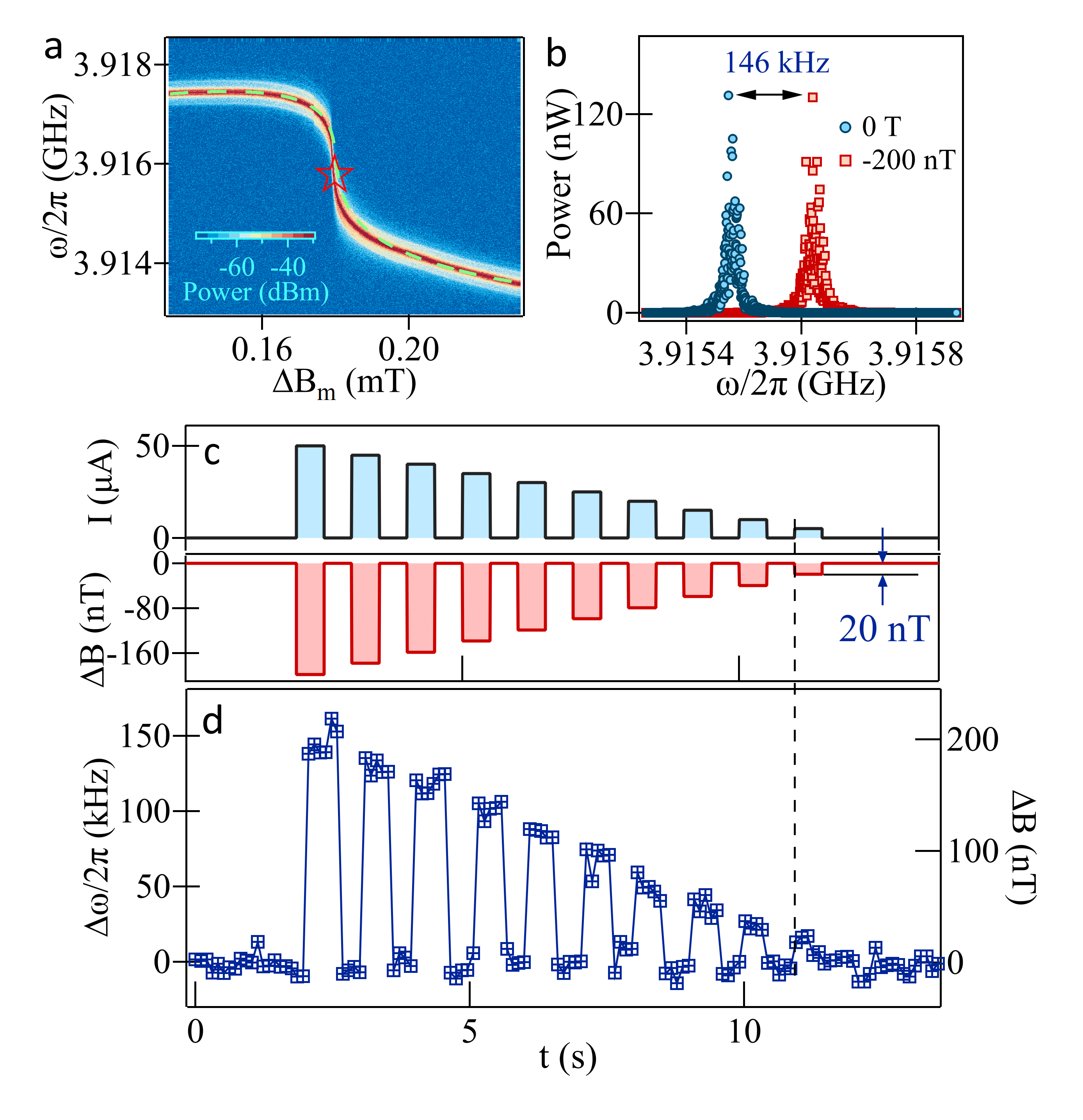, width=8.5cm} \caption{(a), Emission spectra of our magnetometer measured at different magnetic fields, when setting the system at the critical coupling state. Red star marks the CP. Green dashed line is calculated from Eq. (\ref{PMSM}). (b), Two emission spectra measured with the target magnetic field set to 0 nT and -200 nT, respectively. During these measurements, the static bias magnetic field is fixed. (c), Current sequence output from a precision current supply, and the target magnetic field produced by it. (d), Frequency shift of the PMSM induced by the target magnetic field. }
\label{F3}
\end{center}
\end{figure}

The PMSM's dispersion, i.e., the $\omega_s$--$B$ relation, can be well grasped by a Hamiltonian model $\mathcal{H}=\mathcal{H}_a+\hbar(\omega_m-i\alpha_t)\hat{m}^\dag\hat{m}+\hbar\sqrt{\kappa\gamma}e^{i\varphi}(\hat{a}^\dag\hat{m}-\hat{a}\hat{m}^\dag)$, where $\hat{m}$ and $\alpha_t$ are the annihilation operator and total damping rate of the magnon mode, respectively. After reaching the steady oscillation, if we ignore the dynamic damping rate in the gain, the photon and magnon mode oscillate at the same frequency $\omega_s$ with a constant phase difference $\theta$, i.e., $\hat{a}=|\hat{a}|e^{-i\omega_st}$ and $\hat{m}=|\hat{m}|e^{-i(\omega_st+\theta)}$. From the Hamiltonian, we can obtain
\begin{subequations}
\begin{align}
&\frac{\kappa\gamma}{2\alpha_t}[\sin2\theta+\sin2\varphi]+\alpha_t \cot(\theta+\varphi)+\gamma_e\Delta B_m=0 \label{PD} \\
&\omega_s=\omega_p-\frac{\kappa\gamma}{2\alpha_t}[\sin2\theta+\sin2\varphi], \label{PMSM}
\end{align}
\end{subequations}
where $\theta\in\left(-\varphi+\pi,-\varphi+2\pi\right)$. Once $\varphi$ is set, the PMSM's frequency is determined by the dependence of $\theta$ on $\Delta B_m$. Figures \ref{F2} (d)-(f) show the variation of $\theta$ calculated from Eq. (\ref{PD}) at three coupling states. The frequency jump in the coherent photon-magnon coupling arises from the sudden change of $\theta$. These jump points are actually SNBs in nonlinear systems \cite{strogatz2018nonlinear, kuznetsov1998elements}, representing the degeneracy of two equilibria: one stable and one unstable (see solid and dashed lines in Fig. \ref{F2} (d)). We define the field difference between two SNBs as $\Delta B_s$, which varies with $\varphi$. When two SNBs coalesce, a CP of the PMSM occurs \cite{strogatz2018nonlinear}. We extract $\Delta B_s$ from the measurement results at each $\varphi$, and summarize their values in Fig. \ref{F2} (g). Using Eq. (1), we can well reproduce the variation of $\Delta B_s$ with $\varphi$. 

To further illustrate the CP of PMSM, we calculate the evolution of $\omega_s$ in a two-dimensional parameter space defined by $\varphi$ and $\Delta B_m$ using Eq. (\ref{PMSM}) (Fig. \ref{F2} (h)). The red and blue circles indicate the SNBs of the PMSM when the magnetic field is swept upward and downward. Their projections onto the bottom plane are shown as the red and blue solid lines. The area enclosed by these two lines marks the bistable region of the PMSM. When changing $\varphi$, two SNBs gradually coalesce, producing the CPs, at which the PMSM's gyromagnetic ratio is theoretically infinite \cite{SM}, i.e., $\gamma_s=d\omega_s/d B\rightarrow\pm\infty$ (solid black lines in Fig. \ref{F2} (h)). 

\begin{figure*} 
\begin{center}
\epsfig{file=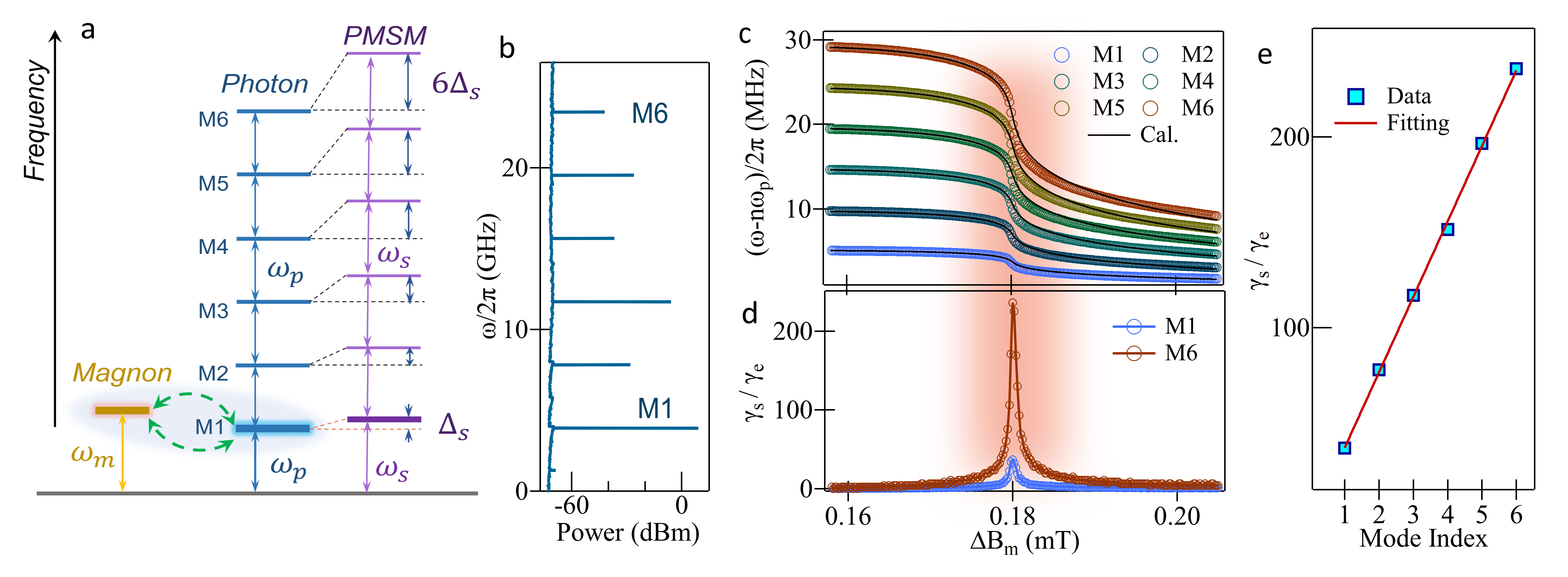, width=17.5cm} \caption{ (a), Schematic of the frequency vernier caliper effect produced by the dynamic damping rate. M1-M6 represent the six orders of the cavity's oscillating modes. (b), Emission spectrum of the active cavity in a wide band. (c), Frequency shift of the PMSM and its high-order oscillating modes. Black solid lines are calculated from Eq. (\ref{PMSM}). (d), Multiples of the gyromagnetic ratio of the PMSM and its six-order oscillating mode ($\gamma_s$) relative to $\gamma_e$. (e), Gyromagnetic ratios of different oscillating modes at the CP following a linear relation of the mode index.}
\label{F4}
\end{center}
\end{figure*}

After realizing CPs of the PMSM, we begin characterizing the PMSM's capability for magnetic field detection. Figure \ref{F3} (a) shows the emission spectra of our magnetometer at different magnetic fields. The red star marks the CP. First, we set the PMSM's frequency at the CP ($\omega_s/2\pi=3.91547$ GHz) by finely adjusting the bias magnetic field. We then send a weak current into the coil to generate a target magnetic field ($\Delta B$) of -200 nT. The PMSM's emission peak shifts to a higher frequency by 146 kHz. Furthermore, we generate a series of target fields by applying a current sequence into the coil. The current amplitudes decrease step-by-step from 50 $\mu$A to 5 $\mu$A with each step lasting 0.5 s. Correspondingly, the target magnetic field can be viewed as a static field, and its strength varies from -200 nT to -20 nT. Figure \ref{F3} (d) shows the frequency shifts of the PMSM induced by the variation in the target magnetic field. The frequency shift occurs within sub-microseconds, demonstrating the instantaneous detection capability of our magnetometer. Due to the high signal-to-noise ratio of the cavity mode, the detection sensitivity of our magnetometer is estimated to be 3.64 fT/$\sqrt{\mathrm{Hz}}$ \cite{forstner2012cavity, xu2024subpicotesla, SM}. Note that our magnetometer is still in the prototype stage, operating at room temperature and in an open environment. Its performance can be significantly improved by reducing noise and further increasing the PMSM's gyromagnetic ratio.

The dynamic damping rate in the saturable gain ($iN\sqrt{\varepsilon/2}\hat{a}$) provides an efficient method to further scale the PMSM's gyromagnetic ratio by multiplying the cavity oscillation to higher frequencies, i.e., $n\omega_p$ with $n = 2, 3, \dots$, as shown in Fig. \ref{F4} (a). Figure \ref{F4} (b) is an emission spectrum of the active cavity, which reveals six orders of oscillating modes. When the fundamental photon mode (M1 in Fig. \ref{F4} (a)) couples with the magnon mode to produce the PMSM, the frequencies of all oscillating modes shift from $n\omega_p$ to $n\omega_s$. The frequency shift of the fundamental photon mode ($\Delta_s=\omega_s-\omega_p$) becomes $n\Delta_s$ for the \textit{n}-th oscillating mode. The small frequency shift of the PMSM is enhanced by the high-order modes, like a ``vernier caliper" in the frequency domain.  

The frequency variation of each oscillating mode of the PMSM is summarized in Fig. \ref{F4} (c). Near the CP ($\Delta B_m\approx 0.18$ mT), the higher the mode index, the sharper the frequency transition. Using Eq. (\ref{PMSM}), we can well reproduce the frequency variation of these modes by multiplying $\omega_s$ with their mode indices. Correspondingly, their gyromagnetic ratios can be calculated from $\gamma_s=nd(\omega_s)/dB$. Figure \ref{F4} (d) shows the multiples of the gyromagnetic ratios of the PMSM and its six-order oscillating mode, relative to $\gamma_e$. The gyromagnetic ratio of the PMSM at the CP is $37\gamma_e$, while this value increases to $236\gamma_e$ for its six-order mode. The measured minimal emission linewidth of the six-order mode is 15 fT, highlighting the excellent magnetic field resolution capability of our magnetometer \cite{SM}. We extract the gyromagnetic ratio of each oscillating mode at the CP and plot them in Fig. \ref{F4} (e). They follow a linear relationship with the mode index, indicating that the gyromagnetic ratio can be further enhanced by increasing the oscillating mode index. 

In conclusion, we utilize the PMSM at CPs to construct a sensitive magnetometer. The PMSM's emission linewidth is reduced to the hertz scale, giving it significantly higher frequency resolution compared to most spin resonance modes. Furthermore, leveraging the sharp frequency transition of the PMSM at the CP, the gyromagnetic ratio of the PMSM is greatly enhanced to $37\gamma_e$, which is further amplified to $236\gamma_e$ by using the sixth-order oscillating mode of the PMSM. As a result, small frequency shifts in the PMSM at the CP become more significant in higher-order oscillating modes. The magnetic field detection sensitivity of our magnetometer is estimated at 3.64 fT/$\sqrt{\mathrm{Hz}}$, demonstrating excellent performance in detecting weak magnetic fields. With appropriate noise suppression techniques, our magnetometer has the potential to achieve sensitivities comparable to or better than those of well-established magnetometers. Our work presents a practical method for weak magnetic field detection and outlines a clear direction for further optimizing detection capabilities. These findings may drive the development of next-generation magnetometers based on cavity magnonics.

\vskip0.25cm 
\begin{acknowledgments}
This work is financially supported by the National Key Research and Development Program of China (Grant No. 2023YFA1406604), the National Natural Science Foundation of China (Grant Nos. 12274260, 12204306, 12304042, and 12474120), STCSM (Nos. 23JC1404100 and 22JC1403300), the Shandong Provincial Natural Science Foundation, China (Grant Nos. ZR2024YQ001 and ZR2024QA187), the Guangdong Basic and Applied Basic Research Foundation (Grant No. 2023A1515110508) and the Qilu Young Scholar Programs of Shandong University.
\end{acknowledgments}

\end{document}